\documentclass[12pt]{article}    
\usepackage{graphicx, cite, epsf, amssymb,epsfig} 
\newsavebox{\LSIM}
\sbox{\LSIM}{\raisebox{-1ex}{$\ \stackrel{\textstyle<}{\sim}\ $}}
\newcommand{\lsim}{\usebox{\LSIM}}
\newsavebox{\GSIM}
\sbox{\GSIM}{\raisebox{-1ex}{$\ \stackrel{\textstyle>}{\sim}\ $}}
\newcommand{\gsim}{\usebox{\GSIM}}
\begin{document} 
%
%
\begin{titlepage} 
\begin{flushright} 
BA-01-50\\ 
DESY 01-201\\ 
hep-ph/0111465 
\end{flushright} 
$\mbox{ }$ 
\vspace{.1cm} 
\begin{center} 
\vspace{.5cm} 
{\bf\large Kaluza-Klein Excitations of W and Z at the LHC?} \\[.3cm]  
\vspace{1cm} 
Stephan J. Huber$^{b,}$\footnote{stephan.huber@desy.de}, 
Chin-Aik Lee$^{a,}$\footnote{jlca@udel.edu} 
and 
Qaisar Shafi$^{a,}$\footnote{shafi@bartol.udel.edu} \\ 
 
\vspace{1cm} {\em  
$^a$Bartol Research Institute, University of Delaware, Newark, USA}\\[.2cm] 
{\em $^b$Deutsches Elektronen-Synchrotron DESY, Hamburg, Germany} 
\end{center} 
\bigskip\noindent 
\vspace{1.cm} 
\begin{abstract} 
Deviations from standard electroweak physics arise in the framework 
of a 
Randall-Sundrum model, with matter and gauge fields in the bulk and the 
Higgs field localized on the TeV brane. We focus in particular on 
modifications associated with the weak mixing angle. Comparison with the 
electroweak precision data yields a rather stringent lower bound of about 
10 TeV on the masses of the lowest Kaluza-Klein excitation of the W and Z 
bosons. With some optimistic assumptions the bound could be lowered to about
7 TeV.
\end{abstract} \end{titlepage} 
 
\section{Introduction}  
The standard model (SM), as formulated for a flat   
four-dimensional space-time, can be easily generalized   
to the RS model \cite{1,CHNOY,GP,DHR,HS2}. In the   
RS model, the four dimensional world   
around us arises from the compactification   
of a curved 5D geometry. The fifth dimension is a  
$S_1/\mathbb{Z}_2$ orbifold and contains two 4D branes at the    
orbifold fixed points at $y=0$ and $y=\pi R$.  
The 5D geometry  is a slice of AdS with the line element  
$ds^2=dy^2+e^{-2\sigma(y)}\eta_{\mu\nu}dx^\mu dx^\nu$,   
where $\sigma(y)=k|y|$, $k$ measures the    
curvature along the fifth dimension, $x$ represents the familiar   
four dimensions and $y$ represents the fifth dimension. The brane at $y=0$ 
is called the Planck-brane   
and the brane at $y=\pi R$ is called the TeV-brane    
because the typical mass scale on each of the branes is of the order   
of the effective 4D Planck mass and a TeV respectively. This exponential 
hierarchy of energy scales is  
generated by the warp factor  $\Omega=e^{-\pi k R}$, with $kR\simeq 11$,  
and provides a new solution to the gauge hierarchy problem.  
  
By compactifying the fifth dimension, a field living in the   
bulk, i.e.~in the full 5D space-time,   
can be decomposed into an infinite number of 4D fields with   
different effective 4D masses using a method known as the    
Kaluza-Klein (KK) decomposition. Depending on the choice of $k$ and $R$,   
there is a large gap between the mass of the    
lightest KK mode  of a bulk field and its next excited state.   
The mass of the first excited state is typically of the    
order of 10 TeV (since experimental constraints rule out   
excited states which are significantly lighter), while the mass    
of the ground state is constrained by experimental data \cite{4,Wa01}, 
since such fields correspond to the ones we    
actually observe in accelerators.   
  
Bulk fields allow one to address problems related to non-renormalizable  
operators within the RS framework, fermion masses 
and mixings   
\cite{GP,HS2}, neutrino masses \cite{GN,HS3} and gauge coupling   
unification \cite{Po,RS01}. With bulk gauge fields, the couplings   
and masses of the weak gauge bosons deviate from their SM values.  
In ref.~\cite{HS1} the modifications of couplings and masses  
were treated independently of each other. This is  
a good approximation if the SM fermions live close  
to one of the two branes. In this letter we present a  
combined analysis of the two effects in order to cover the case 
where the fermions are weakly localized or delocalized  
in the extra dimension as well. Weakly localized fermions are expected for 
instance, if the warped geometry also induces   
the fermion mass hierarchy \cite{HS2}.    
   
We find that the lowest KK excitations of gauge bosons and fermions   
have to be heavier than about 10 TeV. There is no window  
for ``light''  KK states even for delocalized fermions, as was  
the case for the constraints derived in refs.~\cite{GP,DHR}.  
Hence, it is questionable if these excitations can be  
directly studied at the LHC. Indirect evidence for extra  
dimensions, e.g.~rare processes such as $n$--$\bar n$  
oscillations or $\mu \rightarrow e+\gamma$ \cite{K,HS3}, are   
therefore very important. Nevertheless, one may speculate 
if the scenario presented here could account for the small  
deviations of the SM predictions from recent electroweak precision 
data \cite{W01,M01}. This might allow a lowering of the bound to around 7 
TeV.

\section{The Kaluza-Klein reduction}  
With  some slight modifications, the SM is   
easily embedded in a warped geometry.  All of the fields are    
assumed to live in the bulk, except the Higgs field, which is   
confined to the TeV-brane \cite{CHNOY,HS1}. (This can be   
considered as an approximation to the case where the Higgs field lives   
in the bulk but is concentrated exponentially around    
the TeV-brane. The ground state of a scalar field behaves as $e^{4\sigma}$   
and for values of $kR$ around 11, this    
approximation is accurate up to a few permille \cite{HS3}.)   
  
\begin{figure}[t]   
\begin{picture}(100,230)  
\put(95,-40){\epsfxsize7cm \epsffile{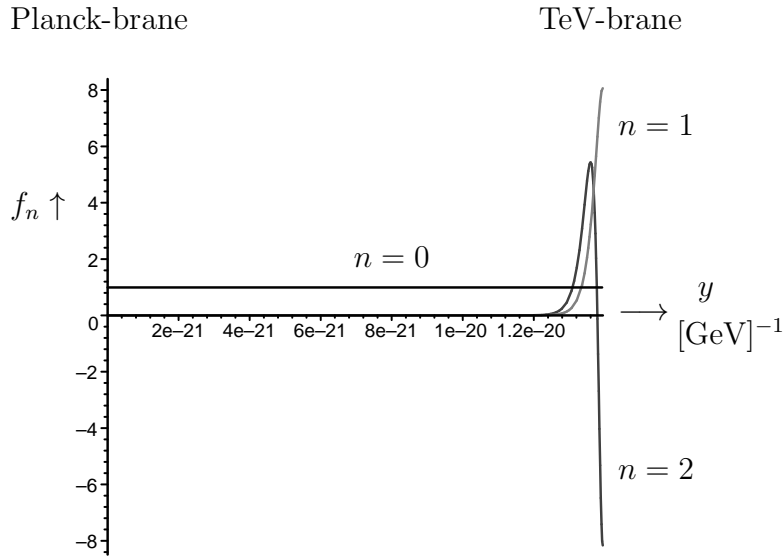}}  
\put(270,210){{TeV-brane}}   
\put(70,210){{Planck-brane}}   
\put(70,140){$f_n\uparrow$} 
\put(300,100){$\longrightarrow$} 
\put(330,109){$y$} 
\put(322,91){[GeV]$^{-1}$} 
\put(200,120){$n=0$} 
\put(300,170){$n=1$} 
\put(300,40){$n=2$} 
\end{picture}   
\caption{The ground state and the first two excited states of the $Z$ boson when   
$kR=10.83$, $k=\overline{M}_{\rm Pl}=2.44\times 10^{18}$ GeV and $a_Z$=0.1849.}  
\label{f_1}  
\end{figure}  
  
The masses   
of the weak gauge bosons are generated by spontaneous    
symmetry breaking arising from the Higgs mechanism,  
$M^2(y)=a^2k\delta(y-\pi R)$, where $a$ is a dimensionless   
parameter which is determined by how strongly the gauge boson   
couples to the Higgs field. Since there is no large hierarchy  
between the weak and KK scales, the 5D mass term   
$M^2$ should be included in the KK reduction of the weak   
gauge bosons from the very beginning \cite {HS1}. The masses  
and 5D wave functions of a gauge field are obtained from its  
5D equations of motion,  
$\frac{1}{\sqrt{-g}}\partial_M(\sqrt{-g}g^{MN}g^{RS}F_{NS})-
M^2(y)g^{RS}A_S=0$. 
Here, the $A_4=0$ gauge is imposed and this is only possible because the 
Higgs field is localized on a brane. If the Higgs were to propagate in the 
bulk instead, it would not be possible to impose both the unitarity and the  
$A_4=0$ gauge simultaneously and we would have to contend with 
$A_4$ 
excitations as well. Using the method of    
separation of variables, a gauge field can be decomposed as follows:    
\begin{equation}A_\mu({x},y)=\frac{1}{\sqrt{2\pi R}}\sum^{\infty}_{n=0}  
A_\mu^{(n)}({x})f_n(y),\label{KK}\end{equation} where    
$A_\mu^{(n)}$ satisfies the field equation for a gauge boson of mass   
$m_n$. This expansion is only valid in the weak coupling limit for 
non-Abelian gauge fields. However, if we drop the requirement that the 
$A_\mu^{(n)}$'s satisfy the 4D equations of motion for a Yang-Mills field, 
any field configuration can be decomposed according to eq. (\ref{KK}). The 
wave functions $f_n$ follow 
from  $(\partial _y^2-2\sigma'\partial_y-M^2+e^{2\sigma}m_n^2)f_n=0$,  
with $\sigma'=d\sigma/dy$. The $f_n$ are normalized by the   
condition $\frac{1}{2\pi R}\int^{\pi R}_{-\pi    
R}f_n(y)^2dy=1$. The solution is given by \cite{DHRP}  
\begin{equation}f_n=  
\frac{e^\sigma(J_1(\frac{m_n}{k}e^\sigma)+b_n    
Y_1(\frac{m_n}{k}e^\sigma))}{N_n},\end{equation}   
where $N_n$   
is the normalization constant, and $b_n$ and $m_n$ are    
obtained by solving the following system of equations \cite{HS1}:    
\begin{eqnarray} 
b(m,a^2)&=&\frac{-\frac{a^2}{2}J_1(\frac{m}{k})+  
\frac{m}{k}J_0(\frac{m}{k})}{-\frac{a^2}{2}Y_1(\frac{m}{k   
})+\frac{m}{k}Y_0(\frac{m}{k})}, \label{3} 
\\[.2cm] 
b_n(e^{-\pi kR}k x_n,0)&=&b(k x_n,-a^2), \nonumber 
\end{eqnarray}   
where $x_n=e^{\pi    
kR} m_n/k$. In fig.~\ref{f_1} we present the wave functions   
of the ground state and the first two excited states of the    
$Z$ boson.  
  
\begin{figure}[t]   
\begin{picture}(100,190)  
\put(60,-50){\epsfxsize8cm \epsffile{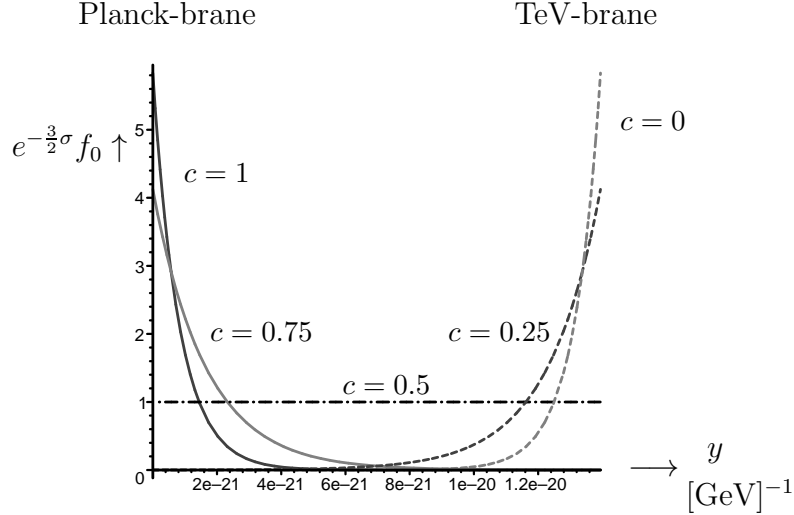}}  
\put(255,180){{TeV-brane}}   
\put(90,180){{Planck-brane}}   
\put(65,130){$e^{-\frac{3}{2}\sigma}f_0\uparrow$} 
\put(298,9){$\longrightarrow$} 
\put(328,16){$y$} 
\put(320,-2){[GeV]$^{-1}$} 
\put(190,40){\small $c=0.5$} 
\put(295,140){\small$c=0$} 
\put(230,60){\small$c=0.25$} 
\put(130,120){\small$c=1$} 
\put(140,60){\small$c=0.75$} 
\end{picture}   
\caption{The wave functions for the ground state fermions for different values of $c$.}  
\label{f_2}  
\end{figure}

Bulk fermions are described by the 5D equation of motion   
$(g^{MN}\gamma_M(\partial_M+\Gamma_M)+m_\Psi)\Psi=0$, where    
$m_\Psi=c\sigma'$ and $\Gamma_M$ is the spin connection in the   
tetrad formulation. The Dirac mass term $m_\Psi$ takes on the functional    
form it has because of the parity restriction $\Psi(-y)=\pm\gamma_5\Psi(y)$.   
For the same reason, the KK    
ground state of a fermion can only be either left-handed or right-handed.   
These ground states are identified with the    
fermion fields actually observed by experiments.   
Again, the 5D field is decomposed into a KK tower \cite{GN,GP},  
\begin{equation}\Psi({x},y)=\frac{1}{\sqrt{2\pi    
R}}\sum^\infty _{n=0}\Psi^{(n)}({x})f^c_n(y),  
\end{equation}   
where $\Psi^{(n)}({x})$ obeys the equation of motion for a   
fermion of mass $m_n$,   
\begin{equation}f^c_0(y)=\frac{e^{(2-c)\sigma}}{N_0},\end{equation}   
and    
\begin{equation}f^c_n(y)=\frac{e^{\frac{5}{2}\sigma}(J_\alpha(\frac{m_n}{k}e^\sigma)+b_n    
Y_\alpha(\frac{m}{k}e^\sigma))}{N_n},\end{equation} for $n>0$.   
The wave functions $f^c_n$ are normalized by   
\begin{equation}\int^{\pi    
R}_{-\pi R}\frac{e^{-3\sigma}f^c_n(y)^2}{2\pi kR}dy=1.\end{equation}   
Fig.~\ref{f_2} shows the wave function for the ground state fermions   
for different values of $c$. For $c>$1/2 ($<$1/2) the zero mode is localized  
towards the Planck-brane (TeV-brane).  For $c$ close to 1/2 the zero mode is
only weakly localized or even delocalized ($c=1/2$).
  
\section{Measuring the weak mixing angle} 
  
Experimentally, there are several ways to measure the weak mixing   
angle, $\theta_W$. Although the equations of the SM remain unchanged    
in 5D, the effective 4D behavior of the fields is slightly modified.   
In particular, measurements of $\theta_W$ made on    
the assumption of a flat 4D space-time would result in a   
value different from the actual 5D weak mixing angle, $\theta_5$.    
The weak mixing angle can be calculated from the ratio   
of the masses of the $W$ and $Z$ bosons,   
\begin{equation}\label{8}  
\cos^2{\theta_1}=\frac{m_W^2}{m_Z^2},   
\end{equation}  
where the subscript 1 indicates that this is the first possible    
definition. Alternatively, $\theta_W$  
can also be obtained from the corresponding ratio of the gauge   
coupling strengths. In the SM, both definitions are equivalent   
(at the tree level).   
  
In the framework of warped models, deviations from standard 4D electroweak  
physics arise.   
The relationship between the (5D) gauge coupling strengths and the   
physical gauge boson mass is no longer linear because of   
eq.~\ref{3} \cite{HS1}. As a result, the $W$ mass gets shifted  
upward compared to the SM.\footnote{  
Curiously, the new experimental value for $m_W$ is indeed   
slightly above the SM prediction \cite{W01}.  
}  
Furthermore, the couplings of the weak gauge bosons  
to fermions are modified. In particular, they become dependent  
on the localization of the fermion within the extra dimension \cite{HS1}.  
The effective 4D coupling between the ground state of a   
fermion and the $n^{th}$ excited state of a gauge boson is    
given by \cite{GP}  
\begin{equation} \label{9}  
g^{(n)}=\frac{g^{(5)}}{(2\pi R)^{3/2}}\int^{\pi R}_{-\pi R} 
e^{-3\sigma}f^c_0(y)^2f_n(y)~dy,  
\end{equation}  
where $g^{(5)}$ is the 5D gauge coupling.

In the approach presented above electroweak symmetry
breaking is treated in the 5D framework.  Corrections to 
weak gauge boson masses and gauge coupling strengths 
arise automatically in the KK reduction. One can interpret
these results also as a mixing effect between a massless
($y$ independent) zero mode of a gauge field and its ($y$
dependent) KK states. 
The mixing between the weak gauge bosons and their KK
excitations is on the order of $m_W^2/M_{KK}^2$. This
explains why all deviations from SM physics are
proportional $1/M_{KK}^2$. Note that somewhat similar 
results are obtained for the case of a flat extra dimension \cite{DPQ}.

In the following, we perform a {\em tree level} comparison  
between the warped model with bulk fields and the ordinary   
SM. This procedure is justified because SM-like quantum   
corrections are essentially the same in both models. The  
tiny modifications (of order $10^{-3}$) of the tree level  
physics will only cause sub-leading corrections in the loops.  
Equivalently, we can just as well work in the picture of a 
gauge field zero mode mixing with its KK states.
In the treatment of the zero mode we then include the usual
SM quantum corrections, while the mixing with the excited
states is approximated at tree-level.  In the literature this
approach has been successfully used to investigate the impact
of extra Z bosons on electroweak observables \cite{EL}, which is
very similar to what is discussed here.
In the warped model (like in models with extra Z bosons), 
there are additional radiative corrections  
coming from loops involving KK states (extra Z bosons). 
However, their influence is small because of the large 
masses (about 10 TeV) of the excited states \cite{EL}.

After electroweak symmetry breaking, the neutral gauge bosons  
in the bulk mix via the 5D weak mixing angle  
$\tan\theta_5=g_1^{(5)}/g_2^{(5)}$, where $g_2^{(5)}$  
and $g_1^{(5)}$ denote the $SU(2)$ and $U(1)$ gauge coupling strengths in  
five dimensions respectively. The gauge coupling strength of the $Z$  
boson in 5D is then given by $g_Z^{(5)}=g_2^{(5)}/\cos \theta_5$.  
Analogous relations hold for the $W$ boson and the photon.   
The effective 4D gauge couplings to fermions, $g_W$, $g_Z$ and  
$g_{\gamma}$ are obtained from  
the $n=0$ case of eq.~\ref{9}, using the relevant 5D gauge  
coupling strengths.  
From these quantities, the weak mixing angle can be defined as  
\begin{equation} \label{10}  
\cos^2\theta_2=\frac{g_W^2}{g_Z^2},  
\end{equation}  
or  
\begin{equation} \label{11}  
\cos^2\theta_3=1-\frac{g_{\gamma}^2}{g_W^2}.  
\end{equation}  
Experimentally, these definitions can be regarded   
as comparisons of the strengths of the charged and neutral  
current, and of the electric and charged current respectively.  
  
In the 4D SM, the values of the weak mixing angle inferred from  
eqs.~\ref{8}, \ref{10} and \ref{11} agree, of course. Since in the   
warped model, the gauge coupling strengths and gauge bosons  
masses are obtained from eqs.~\ref{3} and \ref{9}, this  
is no longer the case.  
The differences between these angles, as measured by   
\begin{equation} \label{deltas} 
\Delta_{ij}=\cos^2\theta_i-\cos^2\theta_j, 
\end{equation} 
 is too small to be    
measured, given the current range of uncertainty in the   
experimental data. However, upper bounds can be set on the    
$\Delta$'s based on the range of uncertainty. This in turn   
provides important constraints on the parameters of the warped SM    
model. Rather than listing the permissible range of values for the 
internal parameters of the model, a more transparent way of expressing 
these constraints   
from an experimental perspective is to   
provide a lower bound for the masses of the first    
KK excitation of the weak gauge bosons.  
  
\begin{figure}[t]   
\begin{picture}(100,210)  
\put(50,-50){\epsfxsize9cm \epsffile{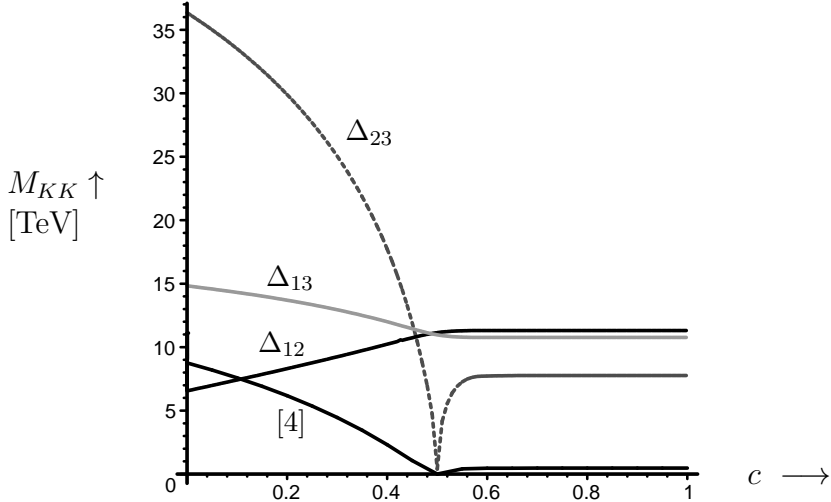}}  
\put(325,10){{$c~\longrightarrow$}}   
\put(45,120){$M_{KK}\uparrow$} 
\put(45,105){[TeV]} 
\put(140,60){{\small $\Delta_{12}$}} 
\put(143,85){{\small $\Delta_{13}$}} 
\put(173,140){{\small $\Delta_{23}$}} 
\put(143,30){{\small \cite{DHR}}} 
\end{picture}   
\caption{The three lower bound constraints and the limit of 
ref.~\cite{DHR} for $k=M_5$ as a function of $c$.}  
\label{f_3}  
\end{figure}  
  
\section{Numerical analysis}  
  
The graphs in this section show the constraints on the mass of the  
first KK excitation for different values of $k/M_5$, where    
$M_5$ (the Planck mass in 5D) has been chosen to fit the observed  
value for the Planck mass in 4D, $M_{\rm Pl}^2\approx M_5^3/k$ \cite{1}.  
In each of the graphs, the lower bounds for the    
mass of the excited $Z$ boson, $M_{KK}\equiv M_Z^{(1)}$ (the 
difference in mass between  
the excited $Z$ and $W$ bosons is insignificant) arising from    
each of the three constraints are plotted.  
The mass of the first    
KK state is adjusted in our model by varying the radius of the  
orbifold. In all the following graphs, we use the experimental    
constraints $\Delta_{12}\leq 1.2\times 10^{-3}$,  
$\Delta_{13}\leq 1.2\times 10^{-3}$ and $\Delta_{23}\leq 1.6\times    
10^{-4}$, based on data from the Particle Data Group \cite{4}.  
The bound on $\Delta_{12,13}$ is dominated by the experimental 
error of the $W$ mass, $M_W=80.419\pm0.056$ GeV, which 
results in an uncertainty in $\cos^2\theta_1$ of $1.1\times 10^{-3}$. 
For the weak mixing angle, we take $\sin^2\theta_W=0.23117\pm0.00016$ with  
the 
error providing the constraint on $\Delta_{23}$.  
 
\begin{figure}[t]   
\begin{picture}(100,190)  
\put(80,-40){\epsfxsize7cm \epsffile{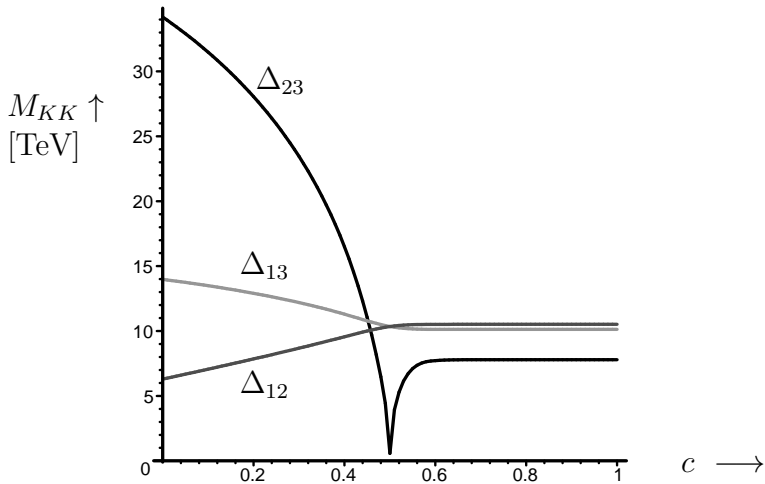}}  
\put(300,5){{$c~\longrightarrow$}}   
\put(45,140){$M_{KK}\uparrow$} 
\put(45,125){[TeV]} 
\put(133,80){{$\Delta_{13}$}} 
\put(133,35){{$\Delta_{12}$}} 
\put(139,150){{$\Delta_{23}$}} 
\end{picture}   
\caption{The three lower bound constraints for $k=M_5/100$ as a function of $c$.}  
\label{f_4}  
\end{figure}  
 
The numerical analysis is done in the following manner. We fix 
the values of $kR$, $k/M_5$ and $c$. From the equation for the 
KK gauge boson spectrum (\ref{3}), we determine the gauge 
boson mass parameters, $a^2(W)$ and $a^2(Z)$ by fitting 
the ground state mass to the experimental $W$ and $Z$  
masses. The masses of the KK excitations are then also 
fixed, as is the value of $\cos^2\theta_1$. 
The 5D weak mixing angle follows from $\cos^2\theta_5=a^2(W)/a^2(Z)$. 
With this information, we can calculate the ratios of the effective 
gauge coupling strengths (\ref{9}) which enter the definitions of 
$\cos^2\theta_{2,3}$. The deviations from SM physics 
are then easily obtained from eq.~(\ref{deltas}). 
We stress again that for different 
values of the KK mass, the deviations scale as 
$\Delta_{ij}\sim1/M_{KK}^2$. 
 
\begin{figure}[t]   
\begin{picture}(100,190)  
\put(80,-40){\epsfxsize7cm \epsffile{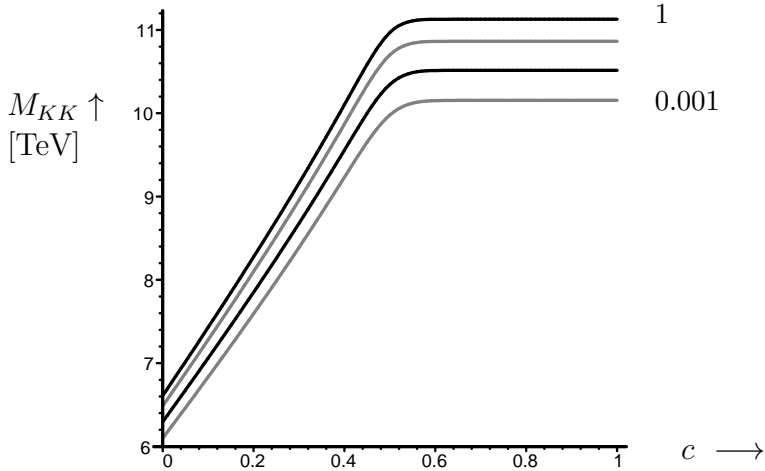}}  
\put(300,10){{$c~\longrightarrow$}}   
\put(45,140){$M_{KK}\uparrow$} 
\put(45,125){[TeV]} 
\put(290,175){\small 1} 
\put(290,142){\small 0.001} 
\end{picture}   
\caption{The constraints obtained from $\Delta_{12}$ for different values of  
$k/M_5=1$, 0.1, 0.01, 0.001 (from top to bottom).}  
\label{f_5}  
\end{figure}  
 
In fig.~\ref{f_3} we display the three different constraints on 
$M_{KK}$ as a function of the 5D fermion mass for $k/M_5=1$. 
For fermions located close to the Planck-brane, $c>1/2$, 
the strongest bound, $M_{KK}\gsim 11.1$ TeV, arises from  
$\Delta_{12}$, i.e.~from the modification of the ratio of the  
weak gauge boson masses. The constraint from $\Delta_{23}$, 
which only involves the effective gauge coupling strengths, gives a 
somewhat lower bound,  $M_{KK}\gsim 7.8$ TeV. 
This is because near the Planck-brane, the gauge boson wave  
function is almost flat and independent of the gauge boson mass  
\cite{HS1}. Therefore, $\Delta_{12}$ and $\Delta_{13}$ lead to 
almost identical results. For  $c\lsim1/2$, the fermions 
become sensitive to the dip in the gauge boson wave function 
at the TeV-brane \cite{HS1} and the gauge couplings deviate 
from their SM values. The constraint from $\Delta_{23}$ takes 
over and pushes the value of $M_{KK}$ far above 10 TeV. For fermions 
strictly confined to the TeV-brane ($c\rightarrow-\infty$), we 
find  $M_{KK}\gsim 62$ TeV. For comparison, we also include the 
lower bound on $M_{KK}$ from  
ref.~\cite{DHR} in fig.~\ref{f_3}.  It arises from the contribution of the 
excited 
gauge bosons to electroweak observables, like for example, effective 
4-fermion 
interactions. Our bounds turn out to be much stronger for 
all values of $c$.\footnote{This 
justifies our neglect of loops containing KK excitations in the 
comparison with the SM.} 
In particular, $M_{KK}\sim 1$ TeV is 
excluded even for fermions close to the Planck-brane. This is the  
main result of the present paper. 

In fig.~\ref{f_4}, we give the corresponding results for $k/M_5=0.01$. 
Qualitatively, the picture is unchanged, but the constraints on $M_{KK}$ 
are slightly looser by about 7 percent. For fermions localized close  
to the Planck-brane, the bound from $\Delta_{12}$ on $M_{KK}$  
is relaxed to 10.3 TeV.  In general, the bound on $M_{KK}$ depends 
logarithmically on $k/M_5$, with the relative change given approximately  
by $0.015\ln(k/M_5)$. In fig.~\ref{f_5}, this behavior is demonstrated 
for $\Delta_{12}$. 

So far we have only included a single 5D mass parameter $c$. If all
SM fermions have a common location in the extra dimension, the 
constraint $c\lsim0.3$ applies. Otherwise the overlap of the fermion
wave functions and the Higgs is insufficient to generate the observed top
quark mass. Moreover, universal values of $c\gsim 1/2$ are disfavored by
deviations from universality \cite{DS}.
Requiring $c\lsim0.3$ for the universal $c$ parameter, the KK scale 
has to be larger than at least 25 TeV,
making this scenario less attractive. However, in the SM there is no 
symmetry that requires the $c$ parameters of different fermion
species to be degenerate. The $c$ parameters can be chosen in such
a way as to generate the fermion masses and mixings without
introducing hierarchies in the 5D Yukawa couplings \cite{HS2}. 
The leptons and light quarks then turn out to have $c$ parameters 
larger than 1/2. Since these particles are the ones used in experiments,
the 11 TeV bound on $M_{KK}$ applies also for this case.    
  
\section{Discussion and summary} 
According to ref.~\cite{DHR}, the LHC can probe $M_{KK}$ up to  
at most 7 TeV. From our results so far, it seems unlikely  
that KK excitations of gauge bosons can be produced directly at  
the LHC.  Still, there may be interesting experimental signatures of the 
presented scenario, like for instance, exotic processes such as 
lepton flavor violation or $n-\bar n$ oscillations \cite{HS3}. 
 
One should keep in mind however, that the crucial input 
in deriving the 10 TeV constraint is the tolerated deviation in the 
$W$ mass. We took $\Delta M_W=56$ MeV \cite{4}. The current  
experimental accuracy on the $W$ mass is about 34 MeV \cite{Wa01}. 
If we were to take this value to be $\Delta M_W$, the quantities 
$\Delta_{12,13}$  
would have to be smaller than $0.86\times10^{-3}$. The lower bound 
on $M_{KK}$ would increase to 13.1 TeV (for $k=M_5$).  
Most interestingly, however, recent data seem to favor a $W$ 
mass  
which is somewhat 
above the SM prediction. At the $1\sigma$ level, the SM prediction  
and the experimental result have no overlap \cite{W01}. The central 
values differ by about 76 MeV for a Higgs mass of 115 GeV. Taking  
this value for $\Delta M_W$ results in $M_{KK}\gsim 9.4$ TeV  
(for $k=M_5$).  For larger Higgs masses, the SM deviates even 
more from experiment. For a Higgs mass of 250 GeV, 
the SM prediction is about 130 MeV lower than the value found  
experimentally. 
This could be accounted for by KK gauge bosons with a mass of 
7 TeV. Thus, being very optimistic, KK gauge boson excitations may  
be at the verge of being discovered at the LHC.

For $M_{KK}\sim7$ TeV, the modifications to the effective gauge coupling 
strengths push $\Delta_{23}$ close to its allowed value, as can be 
observed in fig.~\ref{f_3}. Since $\Delta_{23}$ is rather  
sensitive to the location of fermions, experiments with different  
quarks and/or leptons may actually result in different values for 
$\sin^2\theta_W$.  This may be of some relevance to the  
$3.6 \sigma$ discrepancy in measuring $\sin^2\theta_{\rm eff}^{\rm lept}$ using  
leptonic and hadronic asymmetries \cite{M01}. 

In more elaborate models using extra branes or non-trivial bulk
mass profiles multi-localization of gauge bosons and fermions is possible. 
\cite{multi}. Then exceptionally light KK states can arise
which may escape the bounds we found in our analysis. To avoid
conflict with experiment these states have to decouple from 
SM physics.

In conclusion, our calculations suggest that the KK excitations of W and Z 
within the RS model are expected to be rather heavy, of the 
order of 10 TeV or so. With some caveats, this may be brought down to 
about 7 TeV, which would make their discovery at the LHC much easier. 
  
\section*{Acknowledgements} 
S.J.H. thanks S. Heinemeyer for a useful discussion, the Bartol Research 
Institute for its hospitality during the crucial stages of this work and 
the Alexander von Humboldt Stiftung for a Feodor-Lynen fellowship. 
We thank Tony Gherghetta for useful comments on the draft and for bringing
ref.~\cite{DPQ} to our attention.
 
This work was supported in part by DOE Contract DE-FG02-91ER40626.

\end{document}